\documentclass[12pt]{article}
\usepackage{epsfig}
\begin{document}
\centerline {\Large \bf Density Functional Theory of}
\vskip 1.0 true cm
\centerline {\Large \bf Hard Sphere Condensation Under Gravity}
\vskip 1.0 true cm
\centerline{ \bf Joseph A. Both and Daniel C. Hong }
\vskip 0.2 true cm
\centerline{ Physics, Lewis Laboratory,
Lehigh University, Bethlehem, Pennsylvania 18015}
\vskip 0.3 true cm

\begin{abstract}
The onset of condensation of hard spheres in a gravitational
field is studied using density functional theory. 
In particular, we find that
the local density approximation yields results identical to those obtained
previously using the
kinetic theory [Physica A {\bf 271}, 192, (1999)], 
and a weighted density functional theory gives qualitatively
similar results, namely, that the temperature at which condensation begins
at the bottom scales linearly with weight, diameter, and number of layers of
particles.
\end{abstract}

\section{Introduction}
In a recent paper, one of the authors (DCH) [1] proposed that
hard spheres in the presence of gravitational field, $g$, undergo a 
condensation
transition, and identified the transition temperature, $T_c$, as a function
of external parameters, i.e.:
\begin{equation}
T_c=mgD\mu/\mu_0
\end{equation}
where $m$ and $D$ are the mass and diameter of the hard spheres, $\mu$ is the
dimensionless layer thickness at $T=0$, and $\mu_0$ is a constant that
reflects the particular manner in which a system packs upon condensing.
It was noted that there exists a critical
temperature $T_c$ below which the total number of particles is not conserved.
This is the temperature at which
the density at the bottom layer becomes the close-packed
density.  
Now, since the hard spheres cannot be compressed indefinitely,
if the temperature is lowered below $T_c$,
then the first layer should remain at the close-packed state, while the 
particles at the second layer try to compact themselves and thus crystallize.  
The crystallization then proceeds upward from the bottom layer as the 
temperature is lowered.  This picture was later
confirmed by Molecular Dynamics simulations [2] for mono-disperse
hard spheres and was extended to the segregation of binary mixtures of
hard spheres of different mass and diameters [3].

In the original work [1], the Enskog kinetic equation was used to obtain the
density profile of hard spheres under gravity.  However,
in an attempt to 
solve a highly nonlinear integro-differential kinetic
equation, it was assumed 
that the equilibrium velocity distribution function,
$f({\bf r}, {\bf v})$,  
factorizes into a product of space and velocity dependent parts, i.e. 
$f({\bf r, v})=G({\bf r})\phi({\bf v})$ and further that
the functional form of $\phi({\bf v})$ is
Gaussian, which should be valid for
elastic hard spheres.
The factorization assumption is an equilibrium ansatz, which
states that
the configurational statistics are separated out from the kinetics when
the system is at equilibrium, so that all the equilibrium quantities
can be obtained from the configurational integral of the partition function.
The equilibrium state is then the configuration that minimizes the free
energy.  Therefore, we find it necessary to obtain the results of Ref. [1]
by the variational method.  
Indeed, we will show in this paper that
essentially identical results follow from an application of density functional
theory (DFT) [5] for liquids to the problem.
We will first employ the
simplest form of the density functional
theory known as the local
density approximation (LDA), which
assumes that the
range of inter-particle interaction is much smaller than the typical length
scale on which $\rho(\bf {r})$ varies [6].  We will show that the LDA and the
Enskog theories are in fact identical, so that in both methods, the 
condensation
temperature is defined as the temperature at which the sum rule breaks down.
Next, we will 
analyze the problem with a simple weighted density approximation (WDA) [6-9], 
which takes into account the local variation of the density function.  In this
approximation, the condensation temperature is defined as the temperature
at which the volume density at the bottom layer reaches the maximum allowed
value.  In this approximation, microscopic information is preserved in the
density profile; notably the formation of a crystal shows up in the
density profile as oscillations.  The peak to peak distance of this
oscillation is approximately the particle diameter.
We will
demonstrate that
the results of both analyses present a picture identical to those
presented in Ref. [1]; in particular we will show how the
value $\mu_0$ that appears in Eq. (1) depends on the approximation.

\section{Local Density Approximation}
The essence of the LDA is to assume that the system may be divided into small
pieces of nearly constant density and then to treat each piece as though it
were part of a homogeneous system [6]. Under these assumptions one may
write a free energy functional:
\begin{equation}
F_{LDA}[\rho] = \int d{\bf{r}}\rho({\bf {r}})\psi(\rho({\bf {r}})) +
\int d{\bf{r}} \rho({\bf {r}}) U_{ext}({\bf {r}}),
\end{equation}
where $\psi(\rho({\bf {r}}))$ is the Helmholtz
free energy per particle in the absence
of an external field and $U_{ext}$ is the potential energy per particle
due to an external field such as gravity.  Minimization of this
functional under the global constraint that the number of particles is given
by
\begin{equation}
N = \int_V d{\bf{r}}\rho({\bf {r}})
\end{equation}
should yield the desired density profile.
To be more specific, we define variables for
hard spheres of mass $m$ confined in a $d$ dimensional
volume $V=L^{d-1}H$ with
$L^{d-1}$ being the cross sectional area of a $(d-1)$ dimensional plane and $H$
being the height of the container along which the gravitational
field is acting. 
The Helmholtz free energy per particle consists of two terms,
\begin{equation}
\psi(\rho) = \psi_{id}(\rho) + \psi_{exc}(\rho) ,
\end{equation}
where $\psi_{id}$ is the ideal gas contribution, 
\begin{equation}
\psi_{id}(\rho) = T(\log(\Lambda^3 \rho) - 1).
\end{equation}
Note that $\psi_{id} = -T\log(z^N)/N$ with the single particle partition
function $z=V(2\pi mT)^3/N!$ [10]
and that we have redefined the thermal wavelength 
$\Lambda \equiv (2\pi mT)^{1/2}$.
Next, $\psi_{exc}$ is
the excess contribution to the free energy is due to the configurational
integral coming from the interactions among particles and
is in general written as the integral:
\begin{equation}
\psi_{exc}=T\int_0^\rho\left(\frac{P}{\rho' T}-1\right)\frac{d\rho'}{\rho'},
\end{equation}
The above equation can be derived from the
thermodynamic relation, $P=-(\frac{\partial F}{\partial V})_T$ with
the chain rule: $(\frac{\partial}{\partial V})_T = \frac{\rho^2}{N}(\frac{
\partial}{\partial \rho})_T$.
Note that $P/\rho T -1$ is the virial sum.
Since gravity acts along the vertical direction $z$, $U_{ext} = mgz$,
and the transverse
degrees of freedom can be integrated out to yield the
free energy functional per unit area:
\begin{eqnarray}
\frac{F_{LDA}[\rho]}{A}\equiv \bar F[\rho]& = &
 \int_0^\infty dz\,\rho(z) \psi_{id}(\rho(z)) + \nonumber \\
& & \int_0^\infty dz\, \rho(z) \psi_{exc}(\rho) + \nonumber \\
& & mg\int_0^\infty dz\,\rho(z) z,
\end{eqnarray}
where $A=L^{d-1}$ is the cross sectional area in the $x-y$ plane.
Minimization of the functional under the constraint Eq. (3)
yields an equation for the density profile $\rho$: 
\begin{equation}
\frac{\delta \bar F[\rho]}{\delta \rho}=
T\log(\Lambda^3\rho(z)) + \psi_{exc}(\rho) + \rho \frac {d\psi_{exc}}{d\rho}
+ mgz = \lambda
\end{equation}
where we have introduced a
Lagrange multiplier, $\lambda$.
Defining $\phi \equiv \rho D^3$ and
$\zeta \equiv z/D$, $\lambda$ should be determined by the sum rule:
\begin{equation}
\mu\phi_c
= \int_0^{\phi_0} d\phi\zeta(\phi)=\int_0^\infty d\zeta \phi(\zeta),
\end{equation}
where $\phi_c$ is the close-packed density,  
$\mu$ is the number of layers of particles in the system at $T=0$, and
$\phi_0$ is the density at the bottom layer.
Note that the particular shape of the density profile will depend on the
functional form of the pressure $P$
or, equivalently, on the functional form of the excess free energy, 
$\psi_{exc}$.
One may use the Enskog pressure
for hard disks or a hard sphere equation of state given by a functional form:
\begin{equation}
P= \rho T[1 + \gamma\rho D^d \chi(\rho)],
\end{equation}
where $\chi(\rho)$ is the pair correlation function evaluated at contact ($r =
D$), and
where $\gamma = \frac{\pi}{2}$ when $d=2$ and $\gamma = \frac{2\pi}{3}$ when
$d=3$.  Then
\begin{equation}
\psi_{exc}(\rho) = \int_0^\rho \gamma\rho' D^d \chi(\rho')\frac{d\rho'}{\rho'}.
\end{equation}
Substituting this form of $\psi_{exc}$ into Eq. (8) and taking the derivative
with respect to $z$ generates, in our non-dimensional variables,
the differential equation
\begin{equation}
\frac{d\phi}{d\zeta} + \frac{mgD}{T}\phi = -\gamma\phi\left [\phi\frac{d\chi}{d\zeta}
+ 2\chi\frac{d\phi}{d\zeta}\right ],
\end{equation}
which is
precisely the result obtained in Ref. [1].
Thus the equivalence
between the LDA and Enskog theory has been shown, and the constant
$\mu_0$ that appears in Eq. (1) 
can also be derived by the density functional theory in the local density 
approximation.

To conclude this section we cite some results of the LDA/Enskog theory.
Note that for the liquid phase, the density profile $\phi(\zeta)$ in Eq. (9)
is a monotonically decreasing function of the height $\zeta$ with its 
maximum value at the bottom.  Further, the maximum density $\phi_0$ is a 
function of temperature, too, with the upper bound $\phi_0 \le \phi_c$.
So, the right hand side of Eq. (9) can be written as $f(\phi_0)/\beta$, 
where $\beta = mgD/T$.  The particular form of the function $f(\phi_0)$ 
depends on the approximation.  Ref. [1] (Eq. (15c) and (16c)) gives the 
functional forms of
$\phi_0$ in 2d using the Ree and Hoover correlation function [11]
\begin{eqnarray}
\chi(\phi)& = &\frac{(1-\alpha_1\phi+\alpha_2\phi^2)}{(1-\alpha\phi)^2},\\
\alpha & = & 0.489351 \, \pi/2, \nonumber \\
\alpha_1 & = & 0.196703 \, \pi/2,  \nonumber \\
\alpha_2 & = & 0.006519 \,\pi^2/4. \nonumber
\end{eqnarray}
and in 3d using the Carnahan-Starling equation of state:
\begin{equation}
\frac{P}{\rho T} = \frac{(1+\eta+\eta^2-\eta^3)}{(1-\eta)^3}.
\end{equation}
where we have defined the volume fraction
$\eta = \frac{\pi}{6}D^3\rho = \frac{\pi}{6}\phi$ in 3d and as
$\eta = \frac{\pi}{4}D^3\rho = \frac{\pi}{4}\phi$ in 2d.
For completeness,
we reproduce these.  First, the 2d result:
\begin{eqnarray}
{f(\phi_0)}_{RH}&=& (1+c_2)\phi_0 + \frac{1}{2}c_1\phi_0^2 + \frac{c_3\phi_0}
{(1 - \alpha\phi_0)} \\ \nonumber 
& & - \frac{c_4}{\alpha}\left(\frac{1}{(1 - \alpha\phi_0)}-1\right)
+ \frac{c_4\phi_0}{(1 - \alpha\phi_0)^2}
\end{eqnarray}
with $c_1 = \pi\alpha_2/\alpha^2\approx 0.0855$,
$c_2 = -(\pi/2)(\alpha_1/\alpha^2 - 2\alpha_2/\alpha^3) \approx - 0.710$,
$c_3 = -c_2$, and 
$c_4 = (\pi/2)(1/\alpha - \alpha_1/\alpha^2 + \alpha_2/\alpha^3)\approx 1.278$.
Next, the 3d result:
\begin{equation}
{f(\phi_0)}_{CS} = \phi_0 - \frac{2\phi_0}{(1-\alpha\phi_0)} + 
\frac{2\phi_0}{(1-\alpha\phi_0)^3},
\end{equation}
where in this expression $\alpha = \pi/6$.
Note that in both 2 and 3 dimensions, $f(\phi_0)$ is a monotonically increasing
function of $\phi_0$.  Hence, it has a maximum at $\phi_0 = \phi_c$, the value
of the close-packed density.  Since $\beta$ or equivalently $T$ and the layer 
thickness $\mu$ are arbitrary control parameters, the sum rule, Eq. (9), 
breaks down when $T \le T_c$, where 
\begin{equation}
\mu\phi_c = f_{max}T_c/mgD\equiv \mu_0 T_c/mgD
\end{equation}

While the scaling of the critical temperature displayed in Eq. (1) is 
independent of the particular equation of state used in the calculation,
the maximum value of $f(\phi_0)$, $\mu_0$, depends on the functional form of 
the density profile, or equivalently, the pressure.  Using the two 
approximations above, and taking the maximum densities as 
$\eta_c=\frac{\pi}{6}\sqrt{2} \approx 0.74$ in 3d, and 
$\eta_c=\pi/(2\sqrt{3})\approx 0.91$ in 2d, we find
\begin{eqnarray}
{\mu_0}_{RH}& = &111.31\,\,\,(2d) \nonumber \\
{\mu_0}_{CS}& = &152.34\,\,\,(3d). 
\end{eqnarray}
At the level of Enskog approximation, $\mu_0$ is quite sensitive to the
density at the bottom, $\phi_0$.
At this point, we find it
appropriate to mention the point made by Levin [16]
that reliable information about the fluid-solid coexistence cannot be
obtained by the LDA, because of its inability to
include the density variations in a highly
structured phase (solid).  When the Enskog approximation breaks down, one
has to either abandon the approximation and search for a better one, or 
modify the approximation by removing the unphysical results.  In the original
paper [1], the latter approach was taken, namely based on physical grounds, 
the crystal regime was replaced by a constant average density, a Fermi
rectangle, and the fluid regime was then fit to the Enskog profile, which
is linked to the Fermi rectangle at the liquid-solid interface.  While 
the proportionality constant $\mu_0$ in Eq. (1) obtained this way seems to 
overestimate, and thus while the Enskog equation fails to locate the precise
point of the liquid-solid transition, the prediction of its existence and 
the scaling relation between the critical temperature $T_c$ and 
external parameters (Eq. (1)) seem to remain true.
More elaborate approximations that do take
into account the local variations in the structured phase yield substantially
lower values for $\mu_0$ (see Fig. 2), which are somewhat close to the values
obtained by a Mean Field theory [2].  In order to show the dependence of 
$\mu_0$ on approximation, we also compute in it 3d by the Percus-Yevick 
compressibility form of the equation of state:
\begin{equation}
\frac{P}{\rho T} = \frac{1+\eta+\eta^2}{(1-\eta)^3}
\end{equation}
which yields equally high values for $\mu_0$,
\begin{eqnarray}
{f(\phi_0)}_{PYC}&=&\frac{\phi_0}{(1-\alpha\phi_0)}
 -3\frac{\phi_0}{(1-\alpha\phi_0)^2}+ \\ \nonumber
& &3\frac{\phi_0}{(1-\alpha\phi_0)^3}, \\ \nonumber
{\mu_0}_{PYC}&=& 185.19.
\end{eqnarray}
The slightly different form, namely the virial form
\begin{equation}
\frac{P}{\rho T} = \frac{1+2\eta+3\eta^2}{(1-\eta)^2}.
\end{equation}
yields
\begin{eqnarray}
{f(\phi_0)}_{PYV}&=&3\phi_0-8\frac{\phi_0}{(1-\alpha\phi_0)}+
6\frac{\phi_0}{(1-\alpha\phi_0)^2} \\ \nonumber
{\mu_0}_{PYV}&=& 86.63.
\end{eqnarray}

We further point out that the breakdown of the sum rule is due to the
fact that the pressure has a singularity at $\eta=1$, and thus it has
a {\it finite} value at the close-packed density $\eta_c$, which is 
necessarily less than one [10].  
If one uses the lattice gas pressure [4],
\begin{equation}
P=-T\log(1-\rho),
\end{equation}
which has a singularity at $\rho=1$, then the condensation temperature
is zero, and the density profile is given by the
Fermi function [5]:
\begin{equation}
\rho(z) = 1/(1+\exp( mg(z-\mu)/T))
\end{equation}

\section {Weighted Density Approximation}

The essence of the WDA, as introduced by Tarazona [6,7] and
Curtin and Ashcroft [8] is to recast Eq. (2), the general form of the 
free energy functional, as
\begin{eqnarray}
F_{WDA}[\rho]& = & \int d{\bf{r}}\rho({\bf {r}})\psi_{id}(\rho({\bf {r}})) +
\int d{\bf{r}}\rho({\bf {r}})\psi_{exc}[\rho_w({\bf {r}})]  \nonumber \\ & & + 
\int d{\bf{r}} \rho({\bf {r}}) U_{ext}({\bf {r}}),
\end{eqnarray}
where $\psi_{exc}[\rho_w({\bf {r}})]$ is now a functional of
$\rho({\bf {r}})$,
depending on $\rho({\bf {r}})$ through the weighted average of the density
 given by
\begin{equation}
\rho_w({\bf r} ) = \int d^3{\bf r'}w(|{\bf r}-{\bf r'}|)\rho({\bf r'}),
\end{equation}
where $w(|{\bf r}-{\bf r'}|)$ is an appropriately chosen weighting function.
Following Tarazona [6], we choose
\begin{equation}
\rho_w({\bf r} ) = \frac{3}{4\pi D^3}
\int d^3{\bf r'}\Theta(D-|{\bf r}-{\bf r'}|)\rho({\bf r'}),
\end{equation}
where $\Theta$ is the unit step function, i.e., we replace the local density
$\rho({\bf r})$ with its average over a sphere of radius equal to the particle
diameter $D$.  Because we assume planar symmetry,
i.e., independence in the $x$ and $y$ directions, we may
integrate out the transverse degrees of freedom and write explicitly the
integral above as a one dimensional integral for $z \ge D$:
\begin{equation}
\rho_w(z) = \frac{3}{4D^3}\int_0^\infty dz' \rho(z') (D^2 - (z-z')^2)
\Theta(D-|z-z'|).
\end{equation}
One needs to be careful near the bottom layer $z=0$, namely,
for $0 < z < D$.
In this case, the weighting cannot be done over a sphere a radius $D$, because
of the infinite potential at $z=0$.  We
propose to carry out the weighting over that {\it part} of the sphere of
radius $D$ and centered at $z$ above the $z=0$ plane.  Thus,
the normalization factor, $C$, that is,
the volume over which the integration is performed, is no longer
$C=\int_{-D}^D (\pi(D^2-z'^2)dz'=4\pi D^3/ 3$, 
but instead is $C=\int_{-z}^{D}(\pi(D^2-z'^2)dz'=
\pi[\frac{2}{3}D^3 + D^2 z -
\frac{1}{3} z^3]$.  Hence, for $z < D$,
\begin{eqnarray}
\rho_w(z)& = \frac{1}{[\frac{2}{3}D^3 + D^2 z -
\frac{1}{3} z^3]}&\int_0^\infty dz' \rho(z') (D^2 - (z-z')^2)\times\nonumber\\
& &\Theta(D-|z-z'|).
\end{eqnarray}

As before, we need to extremize the free energy functional under the
global constraint on particle number, so we again use the method of
Lagrange multipliers and
functional differentiation.  Performing the minimization of the
free energy functional (Eq. (7) with $\rho$ in the
excess term replaced by (28)),
we find the following equation must hold:
\begin{eqnarray}
T\ln ({\Lambda}^{3}\rho) &+& \psi_{exc}(\rho_w(z)) +
\int_0^\infty dz' \rho(z')\frac{\delta \psi_{exc}(\rho_w(z'))}
{\delta \rho(z)}\nonumber\\ & &+ mgz +\lambda=0.
\end{eqnarray}
We write explicitly the integral
term in the equation above:
\begin{eqnarray}
\int_0^\infty dz' \rho(z')\frac{\delta \psi_{exc}(\rho_w(z'))}
{\delta \rho(z)}&
= &\int_0^\infty dz'\rho(z')A (z')\nonumber\\
& & \times B (z,z')\Gamma (z'),
\end{eqnarray}
\begin{equation}
A (z') =
\frac{d\psi_{exc}(\rho_w(z'))}{d\rho_w(z')},
\end{equation}
\begin{equation}
B (z,z') =
\left (D^2 - (z - z')^2 \right ) \Theta(D - |z-z'|),
\end{equation}
\begin{equation}
\Gamma (z') = \frac{3}{4D^3}
\end{equation}
if $z' \ge D$ and
\begin{equation}
\Gamma (z') = \frac{1}{[\frac{2}{3}D^3 + D^2 z' -
\frac{1}{3} z'^3]}
\end{equation}
if $0< z' < D$.

The integral equation for $\rho(z)$, Eq. (30), is highly
nonlinear and complex.  Thus, it requires
numerical solution.  We choose to solve Eq. (30) using the Carnahan-Starling
equation of state, Eq. (14), so that
\begin{eqnarray}
A (z')& =&
-\frac{2}{\rho_w(z')\left (1-\frac{\pi}{6}D^3\rho_w(z')\right )}\nonumber\\
& &+ \frac{2}{\rho_w(z')\left (1-\frac{\pi}{6}D^3\rho_w(z')\right )^3}.
\end{eqnarray}
For a given choice of $\lambda$ we iterate Eq. (30) until the iteration
converges to a unique profile.  The integral of the profile (Eq. (9))
determines $\mu$, effectively the depth of the unexcited sample, so for
fixed $m,$ $g$, $D$, and $T$, we tune $\lambda$ to control the number of
particles.
\begin{figure}[tb]
\begin{center}
\epsfig{file=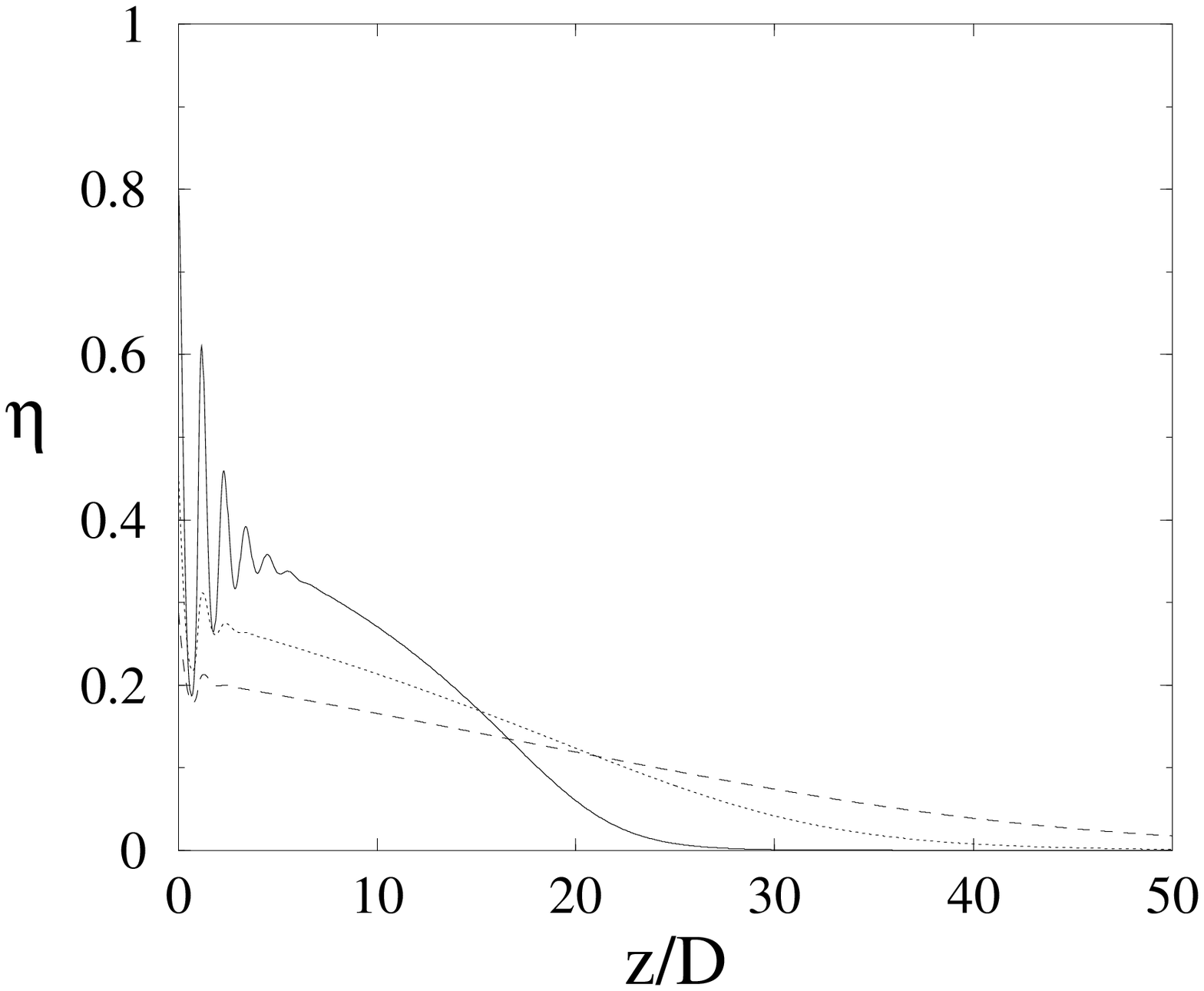,height=5cm,angle=0.,clip=}
\epsfig{file=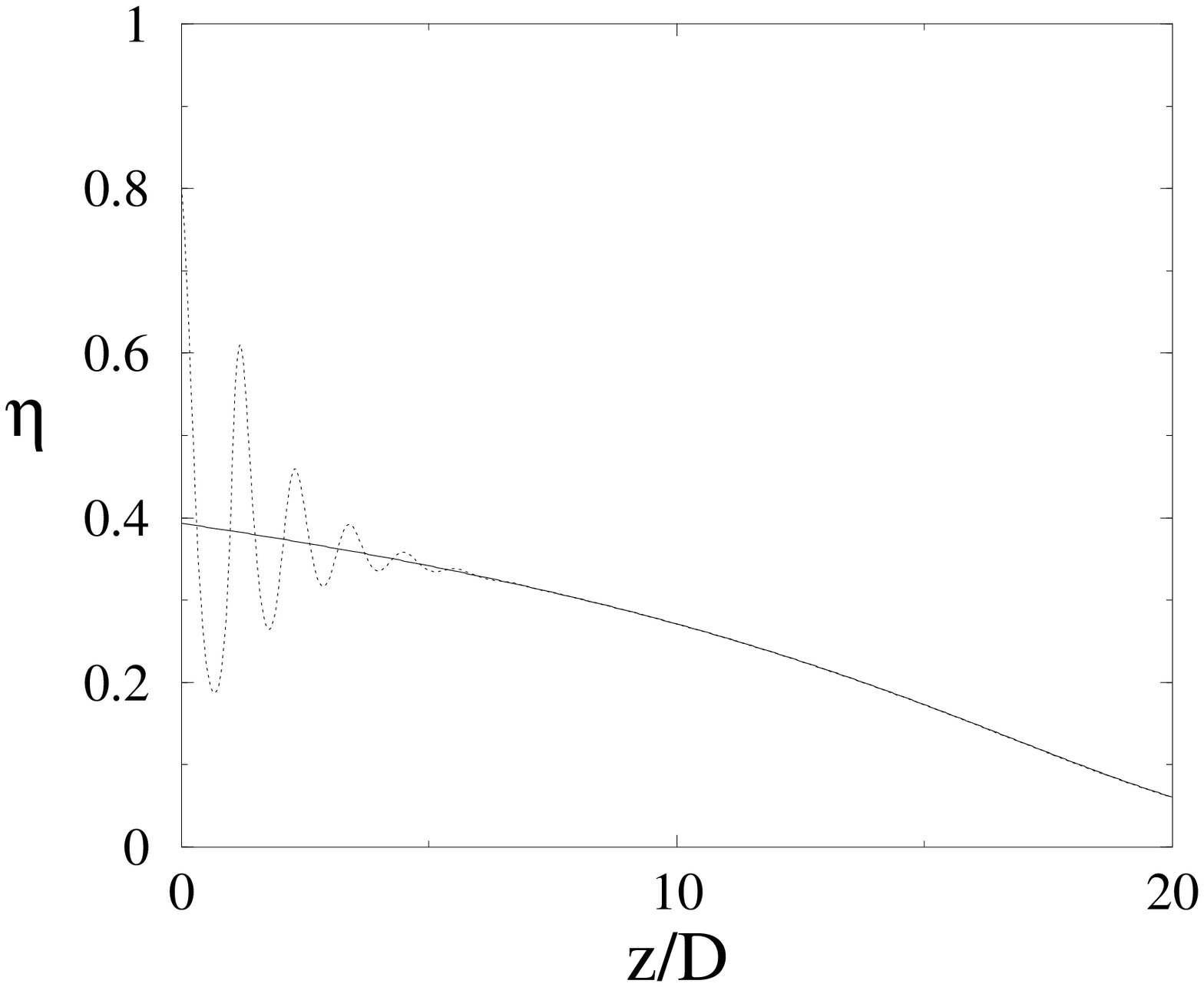,height=5cm,angle=0.,clip=}
\end{center}
\caption{(a) The volume density $\eta$ as
a function of the dimensionless height $z/D$ at different temperatures as 
calculated from numerical solution of Eq. (30) for given set of 
$m$, $g$, $D$, and $\mu$.  The tail extends as the temperature increases.
At high temperature $T$, the density profile is a monotonically 
decreasing function of $z/D$, but as the
$T$ approaches the critical temperature $T_c$,
oscillations develop near the bottom, indicating layer formation.
(b) The dotted line is a magnified view of the topmost curve in Fig. 1a,
while the solid line is the prediction made by the LDA/Enskog theory for the
same system.} 
\end{figure}

\begin{figure}[tb]
\begin{center}
\epsfig{file=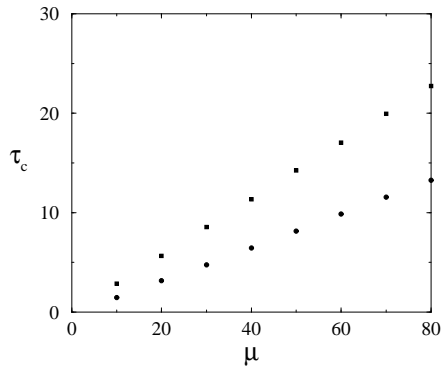,height=5cm,angle=0.,clip=}
\end{center}
\caption{ The dimensionless condensation temperature, $\tau_c\equiv T_c/mgD$, is plotted against the
dimensionless layer thickness $\mu$.  The slope is $1/\mu_0$.  Squares are the
result for $d=2$, circles for $d=3$. }
\end{figure}

We find that at high temperatures, the profiles obtained using the WDA 
match very well
the profiles obtained for the same set of parameters using the LDA/Enskog
approach.  But as we lower the temperature of the system, particles at the
bottom begin to compact themselves, and the crystallization
sets in.  One of the notable features of this weighted density functional
approach is that formation of the crystal can be captured in the density
profile, in particular oscillations in the density profile appear near
$z = 0$.  Fig. 1a shows the 
development of these density peaks for a representative 
system at three different temperatures above $T_c$.  The peak to peak distance
of this oscillation is slightly greater than the diameter of the hard 
sphere.  Fig. 1b is a closer 
view of the density profile for the coolest of these systems (dotted line).
This figure also plots the LDA/Enskog result for the same system
(solid line); it agrees well with 
the WDA profile for large $z/D$, but cannot reflect the rapid oscillations 
in density which occur near the bottom of the sample.

With sufficiently low temperature, the bottom-most peaks in
the profile approach (and even exceed) the physical limit of $\eta = 1$.
Note that the maximum $\eta$ for close-packing in 3d is $0.74$.
However, in our numeric solution, we have chosen the lattice spacing,
$\Delta z = D/256$, with $D$ the particle diameter.  This choice is
to guarantee precision in the solution of the integral equation Eq. (30).
  Hence, even though the physically relevant limit for the volume fraction $\eta$ in 3d 
is 0.74, in our numerical solutions, $\eta$ must approach one at the
close-packed density.  With this modification, we
define the temperature at which $\eta$ first reaches this physical limit 
at $z=0$ as
the critical temperature, $T_c$.
In Fig. 2, for a given set of $m, g, D$, we have plotted the dimensionless 
critical temperature $\tau_c\equiv T_c/mgD$ as a function of the 
initial layer thickness, $\mu$.
The numerically determined value from the slope for the constant $\mu_0$ is
$\mu_0 = 6.10$ in 3d.  We have performed an analogous 
WDA calculation in 2d
using the Ree and Hoover correlation function $\chi(\phi)$, Eq. (13).
The data from this calculation also appear in Fig. 2.  They yield 
$\mu_0 = 3.52$ in 2d.  Both the 2d and 3d WDA results are smaller than
those obtained by the LDA/Enskog approach, and we discuss this next.
\begin{figure}[tb]
\begin{center}
\epsfig{file=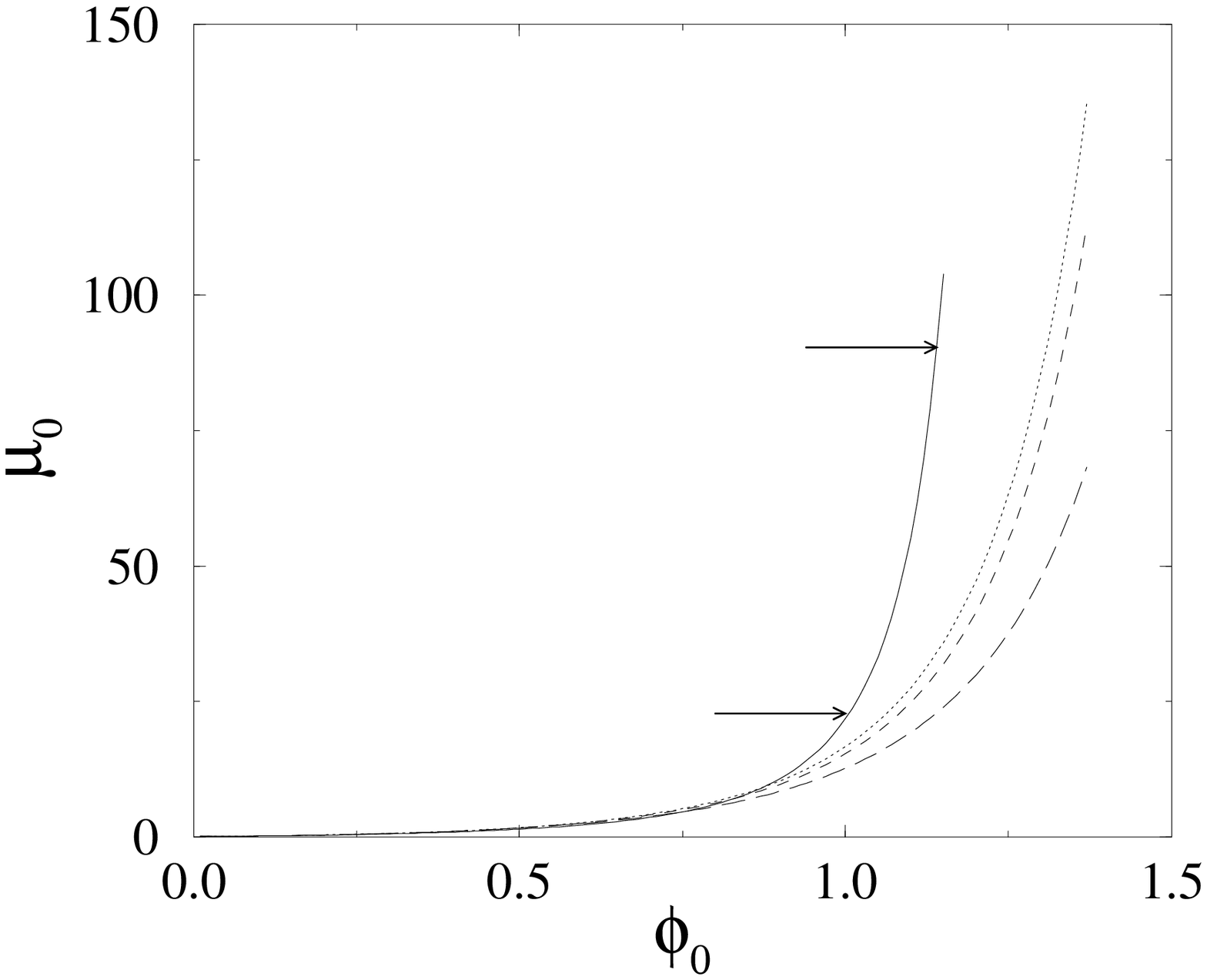,height=5cm,angle=0.,clip=}
\epsfig{file=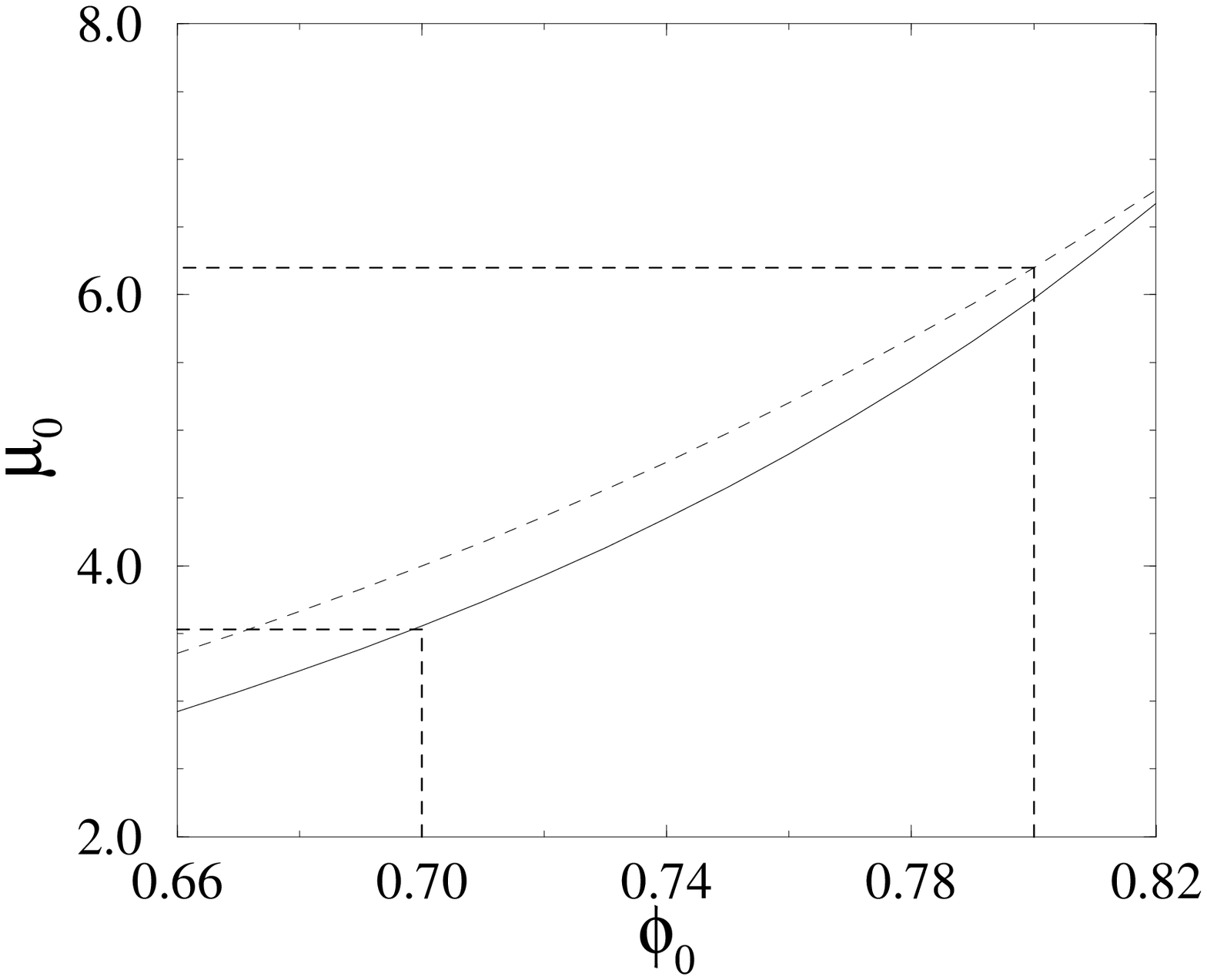,height=5cm,angle=0.,clip=}
\end{center}
\caption{(a) The value $\mu_0\equiv f(\phi_0)$ as a function of density at
 the bottom of 
the sample, $\phi_0$, as calculated in the LDA/Enskog theory.  The solid curve
is for 2d using the Ree and Hoover value of $\chi(\phi)$, Eq. (15).  The 
remaining curves are for 3d: dotted, Percus-Yevick compressibility form, 
Eq. (20); 
dashed, Carnahan-Starling, Eq. (16); long dashed, Percus-Yevick virial form,
Eq. (22).  The arrows on the 2d curve indicate the range of $\mu_0$ calculated
with the LDA/Enskog theory from molecular dynamics simulation values of 
$\phi_0$. (b) An expanded region of Fig. 3a, showing only the 2d Ree and 
Hoover curve (solid) and the 3d Carnahan-Starling curve (dashed).  See the
text for further discussion.} 
\end{figure}

As we have discussed, in the LDA/Enskog approach, the value of $\mu_0$ depends
on $\phi_0$, the density at the bottom, and is identical to the function 
$f(\phi_0)$.  In all the approximations we have used in this work, $f(\phi_0)$ 
(see Eqs. (14, 15, 20, 22)) is a function very sensitive to $\phi_0$ for
$\phi_0$ near close-packed values, i.e. for $\phi_0 \ge 1$.  Fig. 3a 
illustrates this sensitive dependence.
Molecular dynamics simulations in two dimensions [12] have shown that $\phi_0$
at $T < T_c$ varies widely.
For one set of simulations using $10^3$ hard disks with $\mu = 20$, defects in 
packing lead to $\phi_0$ occupying the range $1.00 < \phi_0 < 1.14$, with
higher densities occuring at lower temperatures.
(Note that for square packing in 2d, $\phi_0 = 1$, while for triangular 
packing, $\phi_0=2/\sqrt{3}\approx 1.155$.)  In LDA/Enskog theory, this range 
of 
$\phi_0$ leads to 
$21.76 < \mu_0 < 90.33$ (the arrows in Fig. 3a indicate this range), the
large range due to the sensitivity of $f(\phi_0)$, while the
 WDA theory presented here gives a smaller value, 
$\mu_0 = 3.52$ for the 2d calculation using the same equation of state.  
Though the discrepancy between the WDA and LDA/Enskog 
results seems large, the source of the discrepancy is easy to 
identify.  To do so, we first remark that any density profile derived 
through the WDA for a given system ($m$, $g$, $D$, $\mu$, and $T$) at 
temperature below $T_c$ (WDA) differs from the LDA/Enskog profile for the 
same system appreciably only near the bottom of the sample, where the WDA 
profile exhibits oscillations and the LDA/Enskog profile is a monotonically 
decreasing function bounded between the peaks and troughs of the
WDA profile (see Fig. 1b).  At the temperature when the bottom-most density 
peak in the LDA profile reaches its maximum value 
$\phi_{max} = \frac{4}{\pi}\eta_{max} = \frac{4}{\pi}$, i.e. at $T_c$ (WDA),
the LDA/Enskog profile for the same system will have a much smaller maximum.  
In our work in 2d, at $T_c$ (WDA), the LDA/Enskog profile has 
$\phi_0 = \frac{4}{\pi}\eta_0 = \frac{4}{\pi}\,0.55 \approx 0.70$,
below the square packing value, while in
3d, at $T_c$ (WDA), the LDA/Enskog profile has
$\phi_0 = \frac{6}{\pi}\eta_0 = \frac{6}{\pi}\,0.42 \approx 0.80$, 
below the simple cubic packing value.  If we use these values to compute 
$\mu_0$ in the LDA/Enskog approach as in Fig. 3b, 
we get $\mu_0 = 3.56$ in 2d and
$\mu_0 = 6.20$, consistent with our determination of $\mu_0$ from the slopes
of the lines in Fig. 2.  We see then that $T_c$ (WDA) is in 
general higher than $T_c$ (LDA/Enskog) for the same systems, and this is
reflected in the lower value of $\mu_0$ in the former approach.

Finally, we turn our attention to the question of whether the condensation 
phenomenon we are considering is a phase transition in the thermodynamic 
sense, i.e., whether condensation corresponds to a discontinuity in the first 
or higher derivatives of the free energy with respect to temperature.  We 
address this question by focusing on the gravitational potential energy 
contribution to the free energy, $U_g = mg\int_0^\infty z \rho(z) dz$, 
which is proportional the center of mass $<z> = \int_0^\infty z \rho(z) dz
/\int_0^\infty \rho(z) dz$.  First we show that in the LDA/Enskog theory, 
which is extended to temperatures below $T_c$ by the assumption that the
density in the frozen layers is given by $\phi = \phi_c$ and that the density
above the frozen layers is given by a vertically shifted LDA/Enskog profile 
[1], a kink in the center of mass develops at $T = T_c$, suggesting a 
first order transition.

To do this we recall that the density profile $\phi(\zeta)$ is given by the 
functional form:
\begin{equation}
\beta \zeta=f(\phi)-f(\phi_0) 
\end{equation}
where $\phi_0$ is the density at $\zeta=0$, and $\beta=mgD/T$.  
Above $T_c$,
\begin{eqnarray}
<\zeta(T)> &=& \frac{\int_0^{\infty}\zeta\phi(\zeta)d\zeta}{\int_0^{\infty}
d\zeta\phi(\zeta)} \nonumber \\
&\equiv&\frac{1}{\beta}I_1/I_2
\end{eqnarray}
where
$$I_1=\int_{\phi_0(T)}^{0}[f(\phi)-f(\phi_0)]\phi \frac{df}{d\phi} d\phi$$
$$I_2=\int_{\phi_0(T)}^0 \phi \frac{df}{d\phi} d\phi$$
Now, we note that for $T$ near $T_c$:
\begin{equation}
\phi_0(T) \approx \phi_c -\alpha (T-T_c)
\end{equation}
where $\alpha>0$.
Then for any integrand $G(\phi)$ we can make the following approximation:
\begin{equation}
\int_{\phi_0(T)}^0 G(\phi)d\phi \approx
\int_{\phi_c}^0 G(\phi)d\phi
-\alpha (T-T_c)G(\phi_c).
\end{equation}
Applying this to the above expression to the integral $I_1$ and $I_2$, 
we find that $<\zeta(T)>$ is linear in $T$ with a quadratic correction.

Below
$T_c$, the density profile develops a kink at $\zeta=L$.  For
$\zeta<L$, $\phi(\zeta)=\phi_c$ the close-packed density, and 
for $\zeta>L$, the 
profile is given by the LDA/Enskog profile Eq. (37), and the thickness of
the frozen layer is given by [1] 
\begin{equation}
L=\mu(1-T/T_c).
\end{equation}
We now compute the center of mass $<\zeta(T)>$:
\begin{eqnarray}
<\zeta(T)>&=&\frac{\int_0^{\infty} \zeta\phi(\zeta) d\zeta}{\int_0^{\infty} 
\phi(\zeta)d\zeta} \nonumber \\
&\equiv &\frac{\int_0^L \zeta\phi_c d\zeta + 
\int_L^{\infty}\zeta\phi(\zeta-L)d\zeta}{\mu\phi_c} 
\nonumber \\
& \equiv & \frac{\phi_cL^2/2 + I}{\mu\phi_c}
\end{eqnarray}
where
\begin{eqnarray}
I&=&\int_0^{\infty}\zeta\phi(\zeta)d\zeta + 
L\int_0^{\infty}\phi(\zeta)d\zeta \equiv I_1+LI_2 \nonumber \\
I_2&=&\phi_c(\mu-L).
\end{eqnarray}
Hence,
$$I=\phi_c\mu^2\frac{T}{T_c}(1-\frac{T}{T_c})+ J$$
where 
$$ J=\int_0^{\infty} \zeta\phi(\zeta)d\zeta 
=\int_{\phi_c}^0\zeta(\phi)\phi
\frac{d\zeta(\phi)}{d\phi}d\phi
\equiv \Lambda/\beta^2 \propto T^2$$
and where
$$\Lambda = \int_0^{\phi_c}[f(\phi_c)-f(\phi)]\phi 
\frac{df(\phi)}{d\phi} d\phi.$$
It therefore follows that
\begin{equation}
 <\zeta(T)> = \frac{\mu}{2}  + \lambda_1 T^2
\end{equation}
where
$$\lambda_1=[\frac{1}{\mu}(\frac{1}{mgD})^2(\frac{\Lambda}{\phi_c}-\frac
{\mu_0^2}{2})].$$ The center of mass scales with temperature 
quadratically below $T_c$ but linearly just above $T_c$; thus, there
is a kink in the center of mass and in the gravitational potential energy
contribution to the free energy, giving rise to a first order transition.

The scaling of $<\zeta>$ with $T^2$ below $T_c$ survives a modification of how 
we represent the frozen region.  Suppose, the density in the frozen region is
not represented by a uniform $\phi_c$ but is instead
given by
\begin{equation}
\phi(\zeta)=\sum_i p_i\delta(\zeta-\zeta_i)
\end{equation}
where $\zeta_i$ is the position of the center of hard spheres and $p_i$ is
its peak density in
the i-th row forming a
crystal.   This is a crude way to approximate
the oscillations in the density profile due to the
crystallization.  Then, $I_1$ in Eq. (43) is replaced by:
\begin{equation}
I_1=\int_0^L \zeta\phi(\zeta)d\zeta = \sum_i \zeta_i p_i
\end{equation}
If $p_i=\phi_c$ for all $i$, then,
\begin{eqnarray}
I_1&=&\phi_c\sum_i \zeta_i= \phi_c[1/2 + 3/2 + 5/2 +.... + (2L-1)/2] 
\nonumber \\
&=& \phi_c L^2/2
\end{eqnarray}
which is the same result as that obtained by assuming
the density profile is approximated by
a Fermi rectangle.

\begin{figure}[tb]
\begin{center}
\epsfig{file=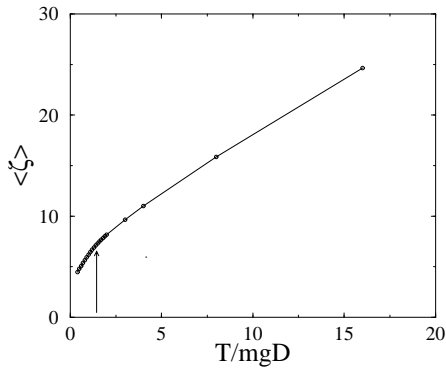,height=5cm,angle=0.,clip=}
\end{center}
\caption{WDA calculation of center of mass $<\zeta>$ vs. $T/mgD$ for a 
system with $\mu = 10$.  The arrow indicates $T_c$ and points to what may be
a kink in the function.} 
\end{figure}

The WDA approach to the problem also yields results suggestive of the existence
of a first order phase transition.  
It may be an artifact of the method that 
in the WDA solutions, as $T$ is lowered
 below $T_c$ as we have defined it, large density peaks whose maxima 
exceed the physical limit $\eta = 1$ appear.  Thus the WDA does not capture 
the exact distribution of material in the system.  However, it is nonetheless
suggestive to examine the dependence of the center of mass $<\zeta(T)>$ on $T$.
Fig. 4 shows results for a representative system in 3d with $\mu=10$ and 
whose critical temperature was found to be on the range 
$1.4\,\, mgD < T_c < 1.5\,\, mgD $.  An elbow, possibly a kink, is apparent in
the vicinity of $T_c$, marking the onset of near linear behavior for $T > T_c$.
We do not assert that this is evidence of a phase transition; we 
display this data merely to suggest that the existence of such a phase 
transition in the WDA approach is not inconsistent with our data.  A different 
form for the weight function in Eq. (26) might yield a better result 
regarding the
nature of the phase transition.

\section{Conclusions}
We have shown that the conclusion of the original paper [1], namely,
that the scaling of the critical temperature at which hard spheres under
gravity begin to form a solid is linear with their weight,
their diameter, and the depth of the sample, necessarily follows from the
simplest density functional theory for the problem (the LDA) and survives a
richer density functional treatment using a WDA.  Prudence requires us to
note that
our WDA for this problem did not include any sophisticated attempt to represent
the crystal-fluid interface, something other researchers [13-15] working on
similar problems have done.  Doing so should likely give a more accurate
quantitative picture than that presented here.

\vskip 0.5 true cm
\noindent {\bf Acknowledgment}
\vskip 0.5 true cm
\noindent The authors wish to thank Y. Levin for helpful discussion 
over the course of this work.
\vskip 0.5 true cm
\noindent {\bf References}
\vskip 0.5 true cm
\noindent [1] D. C. Hong, Physica A {\bf 271}, 192 (1999).

\noindent [2] P. V. Quinn and D. C. Hong, cond-mat/0005196.

\noindent [3] D. C. Hong, P. V. Quinn and S. Luding, cond-mat/0010459.

\noindent [4] D. C. Hong and K. McGouldrick, Physica A {\bf 255}, 415 (1998).

\noindent [5] H. Hayakawa and D. C. Hong, Phys. Rev. Lett. {\bf 78}, 2764 
(1997).

\noindent [6] P. Tarazona, Mol. Phys {\bf 52}, 81 (1984).

\noindent [7] P. Tarazona, Phys. Rev. A {\bf 31}, 2672 (1985).

\noindent [8] W. A. Curtin, N. W. Ashcroft, Phys. Rev. A {\bf 32},
2909, (1985).

\noindent [9] A. Diehl, M. N. Tamashiro, M. C. Barbosa , and Y. Levin, Physica A {\bf 274}, 433 (1999).

\noindent [10] J. P. Hansen and I. R. McDonald, {\it Theory of Simple
Liquids}, Academic Press, New York, 1976.

\noindent [11] F. R. Ree, W. G. Hoover, J. Chem. Phys. {\bf 40},
939, (1964).

\noindent [12] P. V. Quinn, private communication.

\noindent [13] R. Ohnesorge, H. Lowen, and H. Wagner, Phys. Rev. A {\bf 43},
2870 (1991).

\noindent [14] R. Ohnesorge, H. Lowen, and H. Wagner, Phys. Rev. E {\bf 50},
4801 (1994).

\noindent [15] T. Biben, R. Ohnesorge, and H. Lowen, Europhys. Lett. {\bf 28},
665, (1994).

\noindent [16] Y. Levin, Physica A {\bf 287}, 100, (2000).
\end{document}